\documentclass[a4paper,11pt]{article}
\pdfoutput=1 

\usepackage{jheppub}

\usepackage[utf8]{inputenc}

\usepackage{changepage}
\usepackage{tocloft}
\usepackage[T1]{fontenc}
\usepackage[utf8]{inputenc}

\usepackage{graphicx}
\usepackage{floatrow}
\usepackage{subfig}
\renewcommand{\figurename}{Fig.}
\usepackage{transparent}
\usepackage{bbold}
\usepackage{eso-pic}
\usepackage[normalem]{ulem}

\usepackage{tabu}
\usepackage{slashed}
\usepackage{multirow}
\usepackage{booktabs}
\usepackage{float}
\floatstyle{plaintop}
\restylefloat{table}

\usepackage{amsmath}
\usepackage{amssymb}
\usepackage{dsfont}
\usepackage{braket}
\usepackage{mathtools}
\usepackage{mhchem}
\usepackage{cancel}
\usepackage{dsfont}

\DeclarePairedDelimiter\abs{\lvert}{\rvert}
\DeclarePairedDelimiter\norm{\lVert}{\rVert}

\makeatletter
\let\oldabs\abs
\def\abs{\@ifstar{\oldabs}{\oldabs*}}
\let\oldnorm\norm
\def\norm{\@ifstar{\oldnorm}{\oldnorm*}}
\makeatother

\usepackage{fancyhdr}
\usepackage{afterpage}
\usepackage{titlesec}
\titlelabel{\thetitle.\quad}
\usepackage{blindtext}
\usepackage[bottom]{footmisc}

\usepackage{hyperref}
\hypersetup{
    allcolors=black
}

\titlespacing\section{0pt}{12pt plus 4pt minus 2pt}{-\parskip}\relax
\titlespacing\subsection{0pt}{12pt plus 4pt minus 2pt}{-\parskip}\relax

\usepackage{color, soul}
\usepackage{xcolor}
\usepackage{listings}

\definecolor{mBlue}{rgb}{0,0,0.6}
\definecolor{mGray}{rgb}{0.2,0.2,0.2}
\definecolor{mRed}{rgb}{0.6,0,0}
\definecolor{backgroundColour}{rgb}{0.92,0.92,0.92}

\lstdefinestyle{CStyle}{
    backgroundcolor=\color{backgroundColour},   
    commentstyle=\color{mBlue},
    keywordstyle=\color{mRed},
    numberstyle=\tiny\color{mGray},
    stringstyle=\color{magenta},
    basicstyle=\footnotesize,
    breakatwhitespace=false,         
    breaklines=true,                 
    captionpos=b,                    
    keepspaces=true,                 
    numbers=left,                    
    numbersep=5pt,                  
    showspaces=false,                
    showstringspaces=false,
    showtabs=false,                  
    tabsize=2,
    language=C
}

\preprint{IFIC/21-06, 
FTUV/21-0323}

\title{$A_{FB}$ in the SMEFT:  precision $Z$ physics at the LHC}

\author[a]{V\'{i}ctor Bres\'o-Pla,}
\author[b]{Adam Falkowski,}
\author[a]{Mart\'{i}n Gonz\'{a}lez-Alonso,}

\affiliation[a]{Departament de F\'isica Te\`orica, IFIC, Universitat de Val\`encia - CSIC, Apt.  Correus 22085, E-46071 Val\`encia, Spain}
\affiliation[b]{Universit\'{e} Paris-Saclay, CNRS/IN2P3, IJCLab, 91405 Orsay, France}

\abstract{We study the forward-backward asymmetry $A_{FB}$ in $pp\to \ell^+\ell^-$ at the $Z$ peak within the Standard Model Effective Field Theory (SMEFT). We find that this observable provides per mille level constraints on the vertex corrections of the $Z$ boson to quarks, which close a flat direction in the electroweak precision SMEFT fit. Moreover, we show that current $A_{FB}$ data is precise enough so that its inclusion in the fit improves significantly LEP bounds even in simple New Physics setups. This demonstrates that the LHC can compete with and complement LEP when it comes to precision measurements of the $Z$ boson properties.
}

\usepackage{natbib}
\usepackage{graphicx}

\newcommand{\fref}[1]{Fig.~\ref{fig:#1}} 
\newcommand{\eref}[1]{Eq.~\eqref{eq:#1}}

\newcommand{\aref}[1]{Appendix~\ref{app:#1}}
\newcommand{\sref}[1]{Section~\ref{sec:#1}}
\newcommand{\cref}[1]{Chapter~\ref{ch:.#1}}
\newcommand{\tref}[1]{Table~\ref{tab:#1}}

\newcommand{\nn}{\nonumber \\}  
\newcommand{\nnl}{\nonumber \\}

\newcommand{\beq}{\begin{equation}} 
\newcommand{\eeq}{\end{equation}} 
\newcommand{\ba}{\begin{array}}  
\newcommand{\ea}{\end{array}} 
\newcommand{\bea}{\begin{eqnarray}}  
\newcommand{\eea}{\end{eqnarray} }  
\newcommand{\be}{\begin{eqnarray}}  
\newcommand{\ee}{\end{eqnarray} }  
\newcommand{\bal}{\begin{align}}
\newcommand{\eal}{\end{align}}   
\newcommand{\bi}{\begin{itemize}}  
\newcommand{\ei}{\end{itemize}}  
\newcommand{\ben}{\begin{enumerate}}  
\newcommand{\een}{\end{enumerate}}  
\newcommand{\bc}{\begin{center}}
\newcommand{\ec}{\end{center}} 
\newcommand{\bt}{\begin{table}}
\newcommand{\et}{\end{table}}  
\newcommand{\btb}{\begin{tabular}}
\newcommand{\etb}{\end{tabular}}  
\newcommand{\bvec}{\left ( \ba{c}}
\newcommand{\evec}{\ea \right )}

\newcommand{\cO}{{\mathcal O}}

\newcommand{\gev}{\mathrm{GeV}}

\def\hc{{\rm h.c.}} 

\begin{document}

\maketitle

\section{Introduction }\vspace{0.4cm}
\label{sec:intro}

Precision measurements represent a crucial ingredient in the search of physics beyond the Standard Model (SM). It is of the utmost importance to analyze such measurements within general frameworks to minimize the theoretical bias. The Standard Model Effective Field Theory (SMEFT)~\cite{Leung:1984ni,Buchmuller:1985jz} is greatly suited for that purpose, since
its only assumption is the existence of a large gap between the electroweak scale and the masses of particles beyond the SM. The SMEFT represents a robust theoretical approach that is systematically improvable, that benefits from the well-known Effective Field Theory (EFT) machinery and that can be applied to a plethora of New Physics models~\cite{deBlas:2017xtg}. Its application to study the effect of heavy new particles in precision measurements is particularly convenient since these measurements are typically carried out at relatively low energies (at or below the electroweak scale).

The electroweak sector is one of the cornerstones  of the SM, with all interactions fixed by the gauge symmetries, with the only free parameters being the gauge couplings. 
At the end of the previous century, experiments in the LEP collider confirmed the robustness of the SM gauge structure to a remarkable precision.
Arguably, the most important legacy of LEP is a set of electroweak precision measurements of the masses and partial decay widths of the  $Z$ and $W$ bosons~\cite{ALEPH:2005ab,Schael:2013ita}.
The precision program initiated by LEP has exerted enormous influence on the particle physics research. 
On one hand, it  offered historically important guidelines for subsequent discoveries of the remaining SM degrees of freedom: the top quark and the Higgs boson. 
On the other hand, it severely restricted the options for new physics near the electroweak scale. 
In this latter context, the benefits of model-independent characterization of electroweak precision observables were quickly recognized, first in the framework of the oblique parameters~\cite{Peskin:1991sw}, and later in the general EFT framework~\cite{Han:2004az}.   

The Tevatron and the LHC also have a place in this endeavour. First, they allow us to expand the precision program onto the top and the Higgs sectors. 
But even in the field of the classic electroweak observables hadron colliders have a chance to compete, despite having a less clean environment than  LEP. 
This was spectacularly demonstrated by the ATLAS measurement of the $W$ mass~\cite{Aaboud:2017svj}, which now dominates the global average~\cite{Zyla:2020zbs}. Also the total width~\cite{Aaltonen:2007ai,Abazov:2009vs} and certain ratios of leptonic decay widths~\cite{Aaij:2016qqz,Aaboud:2016btc,Aad:2020ayz}  of the $W$ boson are measured more precisely in hadron colliders than in LEP-2. 
A similar result is however lacking regarding the $Z$ boson mass and couplings. 
Naively, this is understandable, as in this case hadron colliders have to compete with a per mille or better accuracy of LEP-1.  
For instance,  
the determinations of the weak mixing angle at the LHC~\cite{Aaij:2015lka,Aad:2015uau,Sirunyan:2018swq} are currently at least a factor of three less precise than the one in LEP. 

In this paper we provide a proof-of-principle demonstration that the LHC precision measurements at the Z-pole can compete with and complement LEP.   
This is the case when electroweak precision measurements are interpreted within the SMEFT framework. 
As we will show, LEP alone cannot simultaneously constrain all higher-dimensional SMEFT operators that modify the $Z$ boson coupling to the up and down quarks.  
Hadron colliders, which probe exactly these couplings when (in particular) the colliding quarks annihilate into an on-shell $Z$-boson, can deliver the missing pieces of information.

Our study will rest on Drell-Yan dilepton production, $pp\to\ell^+\ell^-$.
The LHC collaborations have published precise measurements of the differential cross-sections in this process~\cite{Aaij:2015lka,Aad:2015uau,Aaboud:2016btc,Khachatryan:2016yte,Aaboud:2017ffb,Sirunyan:2018swq}. 
In fact, this is the same process from which the LHC determinations of the SM weak mixing angle are extracted, which, as mentioned above, cannot yet compete with LEP. 
But the  comparison between LEP and LHC is altered once contributions from physics beyond the SM are considered. The fact that LHC measurements of the differential cross-section in $pp\to\ell^+\ell^-$ agree with the SM predictions to such a high precision represents a nontrivial constraint on possible non-standard contributions. The same holds for different extractions of the weak angle, which are in general sensitive to different New Physics  effects. However, the extraction of such bounds cannot be carried out from the results of the SM analysis, but requires a dedicated study.

In the context of the SMEFT, it is known that the study of $pp\to\ell^+\ell^-$ events with high invariant dilepton masses provides strong constraints on 4-fermion contact interactions~\cite{Cirigliano:2012ab,deBlas:2013qqa,Greljo:2017vvb}. This is so because the latter grow with the energy of the process, contrary to the SM contribution, which is suppressed at high energies by the $Z$ propagator. 
However, the situation is quite different for the couplings of the $Z$ boson to fermions ($Zff$). Measurements carried out by LEP1 at the $Z$ peak were able to probe these couplings with very high accuracy. Contrary to the case of 4-fermion interactions, the contributions to the $pp \rightarrow Z/\gamma^* \rightarrow  l^+ l^-$ cross-section from non-standard $Zff$ couplings do not grow with energy. 
Thus one might naively expect that LHC data will not improve significantly our knowledge of the $Zff$ couplings. 

The purpose of this work is to study qualitatively and quantitatively this question in the SMEFT setup. In order to do that, we focus on the forward-backward (FB) asymmetry, $A_{FB}$, which is particularly clean both experimentally and theoretically~\cite{Khachatryan:2016yte,Aaboud:2017ffb}. Contrary to the naive expectation, we find that current $A_{FB}$ measurements provide unique information about the $Z$ couplings to quarks that improve significantly LEP-only bounds. This is true not just in complicated scenarios requiring intricate cancellations between many Wilson Coefficients, but also in simple theory setups.
To illustrate this last point, we anticipate one of our results in~\fref{deltag}, which shows the experimental bounds on possible modifications of the $Z$ couplings to left- and right-handed up quarks, assuming all other interactions are SM-like.
The impact of the inclusion of LHC $A_{FB}$ data is clear.
The reason why LHC can improve on the LEP determination of the $Zuu$ couplings is that the latter suffers from large correlations.

\begin{figure}[b!]
\centering
\includegraphics[scale=0.7]{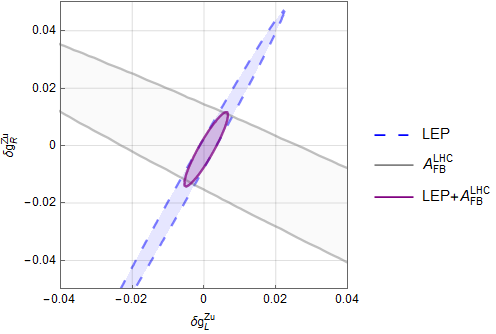}
\caption{Allowed regions (at 95\% CL) for the corrections of the $Z$ couplings to left- and right-handed up quarks. All other couplings are assumed to be given by their SM values.}
\label{fig:deltag}
\end{figure}

The rest of this work is organized as follows. In~\sref{TH} we describe the SMEFT theory setup,  and in~\sref{exp} we review the constraints obtained from ``traditional" pole data (mainly LEP, but not only). ~\sref{hadron} represents the main element of this work. The $pp\to\ell^+\ell^-$ FB asymmetry is introduced and used to extract bounds on the $Zqq$ vertex corrections, which are then compared with LEP bounds. Finally,~\sref{conclusions} contains our conclusions.

\section{Theory framework} \vspace{0.4cm}
\label{sec:TH}

The SMEFT Lagrangian~\cite{Leung:1984ni,Buchmuller:1985jz} is organized into an expansion in $1/\Lambda^2$, where $\Lambda$ is interpreted as the mass scale of new particles in the UV completion of this EFT. 
We truncate the expansion at $\cO(\Lambda^{-2})$, that is, we neglect operators with dimensions higher than six. 
The Lagrangian takes the form
\begin{equation}
\label{eq:TH_Lsmeft}
{\cal L}_{\rm SMEFT} = {\cal L}_{\rm SM}+ {1 \over \Lambda^2} \sum_{i}  c_{i}\, O_{i}^{D=6}, 
\end{equation}
where $ {\cal L}_{\rm SM}$ is the SM Lagrangian, 
each $O_i^{D=6}$ is a gauge-invariant operator of dimension $D$=6, and $c_i$ are the corresponding Wilson coefficients. 
$O_{i}^{D=6}$ span the complete space of dimension-6 operators, see Refs.~\cite{Grzadkowski:2010es,Contino:2013kra} for examples of such sets.  

The SMEFT Lagrangian contains all fundamental interactions predicted in the SM, 
as well as new interactions and deformations of the SM ones due to the dimension-6 operators.  
In this paper we focus on $Z$ and $W$ pole observables, which are mainly sensitive to non-derivative interactions between the heavy electroweak vector bosons and fermions. 
We parametrize them as
\bea
\label{eq:TH_Lvff}
{\cal L}_{\rm SMEFT} & \supset  & 
- {g_L \over \sqrt 2}  \left ( 
W_\mu^+ \bar u_L \gamma_\mu (V + \delta g^{Wq}_L ) d_L 
+  W_\mu^+  \bar u_R \gamma_\mu \delta g^{Wq}_R  d_R  + \hc \right )
\nn
&&- {g_L \over \sqrt 2}  \left ( 
W_\mu^+  \bar \nu_L  \gamma_\mu ( {\bf I} + \delta g^{We}_L)  e_L  
+ \hc \right )
\nn
&&- \sqrt{g_L^2 + g_Y^2} Z_\mu   \left [  
   \sum_{f \in u, d,e,\nu}  \bar f_L  \gamma_\mu ( (T^3_f - s_\theta^2 Q_f) \,{\bf I} +   \delta g^{Zf}_L)  f_L  \right ]\nn
&&- \sqrt{g_L^2 + g_Y^2} Z_\mu   \left [  
   \sum_{f \in u, d,e}   \bar f_R  \gamma_\mu (  - s_\theta^2 Q_f \,{\bf I}  +  \delta g^{Zf}_R) f_R   \right ]. 
\eea
In this Lagrangian, the SM fermion fields $f$ are 3-vectors in the flavor space, written in the basis where their mass terms are diagonal (for neutrinos, where their charged current interactions are diagonal in the limit $\delta g^{We}_L \to 0$), and $V$ is the unitary CKM matrix. 
The SM massive vector fields are denoted as $W_\mu^\pm$ and $Z_\mu$.  
In our conventions, their mass terms take the form
\beq
{\cal L}_{\rm SMEFT} \supset
{g_L^2 v^2 \over 4}\big ( 1 + \delta m_w \big )^2  W_\mu^+ W_\mu^- 
+ {(g_L^2 + g_Y^2)v^2 \over 8} Z_\mu Z_\mu  . 
\eeq 
Here, $g_L$, $g_Y$ are the electroweak couplings, and $v$ is the Higgs VEV. 
These electroweak parameters are assigned numerical values by matching them to the input observables $G_F$, $\alpha(m_Z)$, $m_Z$: 
\beq
\label{eq:TH_couplings}
g_L =  0.6485,  \quad 
 g_Y =  0.3580, \quad 
 v  =  246.22~\gev , 
\eeq  
where the errors can be ignored for the present purpose. 
The input observables are expressed in terms of the electroweak parameters 
as $G_F = (\sqrt 2 v^2)^{-1}$, 
$\alpha(m_Z) = g_L^2 g_Y^2/(4 \pi (g_L^2 + g_Y^2))$, 
$m_Z = \sqrt{g_L^2 + g_Y^2} v/2$, 
i.e. all the corrections due to the dimension-6 operators and loops are absorbed in the definition of $g_L$, $g_Y$, $v$. 
Given the numerical values in \eref{TH_couplings},  the strength of the non-derivative 
$V f \bar f$ interactions is completely fixed in the SM in terms of the fermions' quantum numbers: 
 the weak isospin $T^3_f$, and the electric charge $Q_f$.
As shown in~\eref{TH_Lvff}, deformations of these interactions due to the dimension-6 operators
are parametrized by the {\em vertex corrections} $\delta g$, which are $3\times 3$ matrices in the flavor space and can be flavor-violating.  
They can be expressed as linear combinations of the Wilson coefficients $c_i$ in \eref{TH_Lsmeft} and they are $\cO(\Lambda^{-2})$ in the EFT counting. 
Not all the $\delta g$'s are independent in the SMEFT framework where dimension-8 and higher operators are neglected. 
Expressing $\delta g$ by the Wilson coefficients in any basis one finds the relation
\beq
\label{eq:TH_dgrelations}
\delta g^{Z\nu}_L  = \delta g^{We}_L  +  \delta g^{Z e}_L,  
\qquad 
\delta g^{Wq}_L  =  \delta g^{Zu}_L V -   V\delta g^{Zd}_L~.
\eeq 
In this work we restrict our attention to the flavor-diagonal part of $\delta g$. 
We also approximate $V$ by the unit matrix when it acts on 
$\cO(\Lambda^{-2})$ terms.\footnote{
Going beyond this approximation requires introducing some input scheme for the CKM matrix elements~\cite{Descotes-Genon:2018foz}. } 

The focus of this paper is the new physics effects parametrized by the diagonal elements of $\delta g$ in \eref{TH_Lvff}. 
On the experimental side we restrict ourselves to the so-called {\em pole observables}, where a single  $Z$ or $W$ boson is produced and decays on-shell.
The important feature of this class of observables is that, at leading order, new physics enters {\em only} via $\delta g$.
Other effects of dimension-6 operators enter the pole observables with various suppression factors.
In particular, the contributions from 4-fermion operators are suppressed by   $\Gamma_V/m_V \sim 1/16 \pi^2$~\cite{Han:2004az} or by a loop factor~\cite{Dawson:2019clf} relatively to those of $\delta g$ and are neglected here. 
As for the dipole interactions, $d_f \bar f \sigma_{\mu \nu} f V_{\mu\nu}$, their interference with the SM amplitudes is suppressed by the small fermion masses, 
while the quadratic effects in $d_i$ are $\cO(\Lambda^{-4})$ in the EFT expansion and are consistently neglected.    
Thus, restricting to pole observables largely simplifies the analysis, allowing one to avoid dealing with the huge parameter space of the SMEFT and instead to focus on a relatively small number of parameters $\delta g$.  
To be specific, we will discuss model-independent constraints on the following 20 independent real parameters: 
\beq
\label{eq:TH_dg}
\delta g^{We}_{L}, \,  \delta g^{W\mu}_{L}, \,  \delta g^{W\tau}_{L}, \,  
\delta g^{Ze}_{L/R}, \,  \delta g^{Z\mu}_{L/R}, \,  \delta g^{Z\tau}_{L/R}, \, 
\delta g^{Zd}_{L/R}, \, \delta g^{Zs}_{L/R}, \,  \delta g^{Zb}_{L/R}, \,
\delta g^{Zu}_{L/R}, \, \delta g^{Zc}_{L/R},  \,  \delta m_w .
\eeq 
As mentioned earlier, in this paper we do not consider observables targeting flavor-off-diagonal elements of the $\delta g$ matrices. 
Therefore all $\delta g$'s above refer to the diagonal elements, 
for which we will employ the shorter notation: 
$ \delta g^{V f_J}_X \equiv [\delta g^{Vf}_X]_{JJ}$.
We also do not consider top observables, 
therefore we do not provide constraints on $\delta g^{Zt}_{L/R}$. 
As for $\delta g^{Wq}_R$, they affect the pole observables quadratically (because there are no SM right-handed currents to interfere with) and thus are neglected as $\cO(\Lambda^{-4})$ effects.
Finally, $\delta g^{Z\nu}_L$ and $\delta g^{Wq}_L$ can be expressed by the vertex corrections in \eref{TH_dg} by virtue of \eref{TH_dgrelations}. 

We will present our results in the form of a likelihood function for the parameters in \eref{TH_dg}, which encodes simultaneous constraints on the 19 $\delta g$'s and $\delta m_w$ including all correlations. 
This can be translated into constraints on the Wilson coefficients in the reader's favorite basis, provided the map between $\delta g$ and the Wilson coefficients is known.\footnote{Alternatively, one can use the Higgs basis defined in Ref.~\cite{deFlorian:2016spz}, 
where $\delta g$'s play the role of the Wilson coefficients and no translation is needed.
}  
In \aref{warsaw} we provide the map relating $\delta g$ to the Wilson coefficients of the commonly used Warsaw basis~\cite{Grzadkowski:2010es}.

\section{Traditional pole observables} \vspace{0.4cm}
\label{sec:exp}

The traditional $Z$-pole observables are those carried out at LEP-1 and SLC, which we list in~\tref{EWPT_zpole}. The experimental values are taken from the classic report of Ref.~\cite{ALEPH:2005ab}, with minor modifications affecting the $Z$ width, the hadronic cross section $\sigma_{had}$ and the FB asymmetry of b quarks to take into account the recent results of Refs.~\cite{Janot:2019oyi,dEnterria:2020cgt}.
We also include the measurements of $A_s$ and $R_{uc}$~\cite{Abe:2000uc,Zyla:2020zbs} that, despite being less precise, are needed to remove flat directions in the EFT parameter space.

In the case of the $W$ boson, the traditional pole observables were carried out at LEP-2, Tevatron and LHC, and they are summarized in~\tref{EWPT_wpole}. We stress that this dataset includes recent LHC results concerning the leptonic branching ratios of the $W$ boson~\cite{Aaij:2016qqz,Abbott:1999pk,Aad:2020ayz}.
Once again, we include the not-so-precise measurements of $R_{Wc}$~\cite{Zyla:2020zbs} to remove flat directions.

\begin{table}
 \begin{center}
 \begin{tabular}{|c|r|c|c|c|}
\hline
{\color{blue}{Observable}} & {\color{blue}{Experimental value}}     
&  {\color{blue}{SM prediction}}    &  {\color{blue} Definition}
  \\  \hline   
$\Gamma_{Z}$ [GeV]  & $2.4955 \pm 0.0023$~ \cite{ALEPH:2005ab,Janot:2019oyi} & $ 2.4941$    & $\sum_f \Gamma (Z \to f \bar f)$
 \\  \hline
$\sigma_{\rm had}$ [nb]  & $41.4802\pm 0.0325$ ~\cite{ALEPH:2005ab,Janot:2019oyi} &  $41.4842$ &  ${12 \pi \over m_Z^2} {\Gamma (Z \to e^+ e^-) \Gamma (Z \to q \bar q) \over \Gamma_Z^2}$
  \\  
 $R_{e}$  & $20.804\pm 0.050$~ \cite{ALEPH:2005ab} &  $20.734$  &    $ {\sum_{q} \Gamma(Z \to q \bar q) \over  \Gamma(Z \to e^+ e^-)} $
   \\  
 $R_{\mu}$  & $20.785 \pm 0.033$~ \cite{ALEPH:2005ab} &  $20.734$  &    $ {\sum_{q} \Gamma(Z \to q \bar q) \over  \Gamma(Z \to \mu^+ \mu^-)} $
   \\ 
 $R_{\tau}$  & $20.764\pm 0.045$~ \cite{ALEPH:2005ab} &  $20.781$  &    $ {\sum_{q} \Gamma(Z \to q \bar q) \over  \Gamma(Z \to \tau^+ \tau^-)} $
 \\  
 $A_{\rm FB}^{0,e}$ & $0.0145\pm 0.0025$ ~\cite{ALEPH:2005ab} &  $0.0162$   &  ${3 \over 4} A_e^2$
  \\  
 $A_{\rm FB}^{0,\mu}$ & $0.0169\pm 0.0013$ ~\cite{ALEPH:2005ab} &  $0.0162$   &  ${3 \over 4} A_e A_\mu $
  \\  
 $A_{\rm FB}^{0,\tau}$ & $0.0188\pm 0.0017$ ~\cite{ALEPH:2005ab} &  $0.0162$   &  ${3 \over 4} A_e A_\tau$
  \\ \hline 
$R_b$ & $0.21629\pm0.00066$~ \cite{ALEPH:2005ab} & $0.21581$  &     ${ \Gamma(Z \to b \bar b) \over \sum_q \Gamma(Z \to q \bar q)}$
   \\  
$R_c$ & $0.1721\pm0.0030$ ~ \cite{ALEPH:2005ab}  & $0.17222$   & ${ \Gamma(Z \to c \bar c) \over \sum_q \Gamma(Z \to q \bar q)} $
\\  
$A_{b}^{\rm FB}$ & $0.0996\pm 0.0016$~ \cite{ALEPH:2005ab,dEnterria:2020cgt}  & $0.1032$  & ${3 \over 4} A_e A_b$
 \\  
 $A_{c}^{\rm FB}$ & $0.0707\pm 0.0035$ ~ \cite{ALEPH:2005ab} &  $0.0736$  & ${3 \over 4} A_e A_c$
  \\ \hline 
 $A_e$ & $0.1516 \pm 0.0021$ ~\cite{ALEPH:2005ab} &  $0.1470$ &  ${\Gamma(Z \to e_L^+ e_L^-) - \Gamma(Z \to e_R^+ e_R^-) \over \Gamma(Z \to e^+  e^-) }$
 \\ 
  $A_\mu$ & $0.142 \pm 0.015$ ~\cite{ALEPH:2005ab} &  $0.1470$ &  ${\Gamma(Z \to \mu_L^+ \mu_L^-) - \Gamma(Z \to \mu_R^+ \mu_R^-) \over \Gamma(Z \to \mu^+  \mu^-) }$
 \\ 
 $A_\tau$ & $0.136 \pm 0.015$ ~\cite{ALEPH:2005ab} &  $0.1470$ &  ${\Gamma(Z \to \tau_L^+ \tau_L^-) - \Gamma(Z \to \tau_R^+ \tau_R^-) \over \Gamma(Z \to \tau^+  \tau^-) }$
 \\ \hline 
  $A_e$ & $0.1498 \pm 0.0049$~ \cite{ALEPH:2005ab} &  $0.1470$ &  ${\Gamma(Z \to e_L^+ e_L^-) - \Gamma(Z \to e_R^+ e_R^-) \over \Gamma(Z \to e^+  e^-) }$
 \\ \hline 
 $A_\tau$ & $0.1439 \pm  0.0043$~ \cite{ALEPH:2005ab} &  $0.1470$ &  ${\Gamma(Z \to \tau_L^+ \tau_L^-) - \Gamma(Z \to \tau_R^+ \tau_R^-) \over \Gamma(Z \to \tau^+  \tau^-) }$
 \\ \hline 
 $A_b$ & $0.923\pm 0.020$~ \cite{ALEPH:2005ab} & $0.935$  &
 ${ \Gamma(Z \to b_L \bar b_L)  -   \Gamma(Z \to b_R \bar b_R)  \over  \Gamma(Z \to b \bar b)   }$
 \\  
$A_c$ & $0.670 \pm 0.027$~ \cite{ALEPH:2005ab} & $0.668$
 &  ${ \Gamma(Z \to c_L \bar c_L)  -   \Gamma(Z \to c_R \bar c_R)  \over  \Gamma(Z \to c \bar c)   }$
 \\ \hline 
$A_s$  & $0.895 \pm 0.091$~ \cite{Abe:2000uc} & $0.936$
&  ${ \Gamma(Z \to s_L \bar s_L)  -   \Gamma(Z \to s_R \bar s_R)  \over  \Gamma(Z \to s\bar s)   }$
\\ \hline
$R_{uc}$ & $0.166 \pm 0.009$ ~\cite{Zyla:2020zbs}  & $0.1722$   & ${ \Gamma(Z \to u \bar u) +  \Gamma(Z \to c \bar c) \over 2 \sum_q \Gamma(Z \to q \bar q)} $
\\ \hline 
\end{tabular}
\end{center}
\caption{
{\bf $Z$  pole observables}. The experimental errors of the observables not separated by horizontal lines are correlated, which is taken into account in the fit. The first $A_e$ and $A_\tau$ values come from the combination of leptonic polarization and left-right asymmetry measurements at SLD, while the second values come from the LEP-1 measurements of the  polarization of the final leptons.
}
\label{tab:EWPT_zpole}
 \end{table}

\begin{table}
 \begin{center}
 \begin{tabular}{|c|r|c|c|c|}
 \hline
{\color{blue}{Observable}} & {\color{blue}{Experimental value}}   
&  {\color{blue}{SM prediction}}    
 \\  \hline  \hline
$m_{W}$ [GeV]  & $80.379 \pm 0.012$ ~\cite{Zyla:2020zbs}    &  $80.356$   
\\ \hline  
$\Gamma_{W}$ [GeV]  & $ 2.085 \pm 0.042$  ~\cite{Zyla:2020zbs} &  $2.088$  
\\ \hline 
${\rm Br} (W \to e \nu)$ & $ 0.1071 \pm 0.0016$ ~\cite{Schael:2013ita} & $0.1082$ 
\\ 
${\rm Br} (W \to \mu \nu)$ & $ 0.1063 \pm 0.0015$ ~\cite{Schael:2013ita} &  $0.1082$  
\\ 
${\rm Br} (W \to \tau \nu)$ & $ 0.1138 \pm 0.0021$ ~\cite{Schael:2013ita} & $0.1081$ 
 \\ \hline 
${\rm Br} (W \to \mu \nu) / {\rm Br} (W \to e \nu)$ 
&   $0.982 \pm 0.024$~\cite{Abulencia:2005ix} 
&  $1.000$  
 \\ \hline 
${\rm Br} (W \to \mu \nu) / {\rm Br} (W \to e \nu)$ 
&   $1.020 \pm 0.019$~\cite{Aaij:2016qqz} 
&  $1.000$  
 \\ \hline 
${\rm Br} (W \to \mu \nu) / {\rm Br} (W \to e \nu)$ 
&  $1.003 \pm 0.010$~\cite{Aaboud:2016btc}
&  $1.000$ 
 \\ \hline 
${\rm Br} (W \to \tau \nu) / {\rm Br} (W \to e \nu)$ &
$ 0.961 \pm 0.061$~\cite{Abbott:1999pk,Zyla:2020zbs} &  $0.999$ 
 \\ \hline 
${\rm Br} (W \to \tau \nu) / {\rm Br} (W \to \mu \nu)$ & $ 0.992 \pm 0.013$ ~\cite{Aad:2020ayz} & $0.999$  
\\ \hline 
$R_{Wc}\equiv { \Gamma(W \to c s) \over  \Gamma(W \to u d) + \Gamma(W \to c s) }$ & $ 0.49 \pm 0.04$ ~\cite{Zyla:2020zbs}  &  $0.50$ 
  \\ \hline
\end{tabular}
\end{center}
\caption{{\bf $W$ pole observables}. 
The experimental errors of the observables not separated by horizontal lines are correlated, which is taken into account in the fit. 
}
\label{tab:EWPT_wpole}
 \end{table}

We use these pole observables to constrain the EFT coefficients in \eref{TH_dg}. For that we build the associated $\chi^2$ function
\beq
\chi^2 = \sum_{ij}\left [ O_{i,\rm exp}  -  O_{i,\rm th} \right ] (\sigma^{-2})_{ij}  \left [ O_{j,\rm exp}  -  O_{j,\rm th} \right ],
\eeq
where  $\sigma^{-2}$ is the inverse of the covariance matrix, and the theory prediction of each observable is given by
\beq
\label{eq:Oith}
O_{i,\rm th}  = O_{i,\rm SM}  +  \sum_k \alpha_{ik}\, \delta g_k~,
\eeq
where $\delta g_k$ runs over all the couplings listed in \eref{TH_dg}. For the SM predictions of the $Z$ observables in  \tref{EWPT_zpole} we use the values obtained in Ref.~\cite{Falkowski:2019hvp}, which are calculated using the couplings in~\eref{TH_couplings} and the known state-of-the-art expressions.
For the $W$ observables in  \tref{EWPT_wpole} we use the SM predictions from Ref.~\cite{dEnterria:2020cpv}, except for the $W$ mass where we use the semi-analytic expression from Ref.~\cite{Awramik:2003rn}. 
The uncertainties of the SM predictions are subleading compared to  the experimental uncertainties and will be neglected in this work.
In our analysis we use the $\alpha_{ik}$ coefficients in \eref{Oith} calculated at tree-level. 
Log-enhanced one-loop corrections generated through running to higher/lower scales can be included via the renormalization group equation~\cite{Alonso:2013hga}. 
Finite loop corrections~\cite{Dawson:2019clf} are neglected in this work; see e.g.~\cite{Dawson:2020oco} for estimates of their effects in the context of a different fit.
We minimize the $\chi^2$ function with all $\delta g$ corrections and $\delta m_w$ present simultaneously. Although all the 20 coefficients are kept all the time (and their bounds are correlated), it is instructive to discuss the results by groups. For the leptonic couplings, we find the following central values and 1$\sigma$ errors:

\bea
\label{eq:dgl}
&&[\delta g^{We}_L]_{ii} = \bvec  
-1.3 \pm 3.2 \\   
-2.8  \pm 2.6 \\  
1.5 \pm 4.0 
\evec \times 10^{-3}, \\
\label{eq:dgZl}
&& [\delta g^{Ze}_L]_{ii} = \bvec  
-0.19 \pm 0.28 \\   
0.1  \pm 1.2 \\  
-0.09 \pm 0.59  
\evec \times 10^{-3},
\quad [\delta g^{Ze}_R]_{ii} = \bvec  
-0.43 \pm 0.27 \\   
0.0  \pm 1.4 \\  
0.62 \pm 0.62 
\evec \times 10^{-3}~.
\eea

For the couplings involving strange, charm and bottom quarks, we obtain

\bea
\label{eq:dgZs}
&&
\delta  g^{Zs}_L = (1.3 \pm 4.1) \times 10^{-2},
\quad\quad 
\delta  g^{Zs}_R = (2.2 \pm 5.6) \times 10^{-2}, \\
\label{eq:dgZc}&&
\delta  g^{Zc}_L = (-1.3 \pm 3.7) \times 10^{-3},
\quad\quad 
\delta  g^{Zc}_R = (-3.2 \pm 5.4) \times 10^{-3}, \\
\label{eq:dgZb}&&
\delta  g^{Zb}_L = (3.1\pm 1.7) \times 10^{-3},
\quad\quad 
\delta  g^{Zb}_R = (21.8 \pm 8.8) \times 10^{-3}~.
\eea
The data also constrain the SMEFT corrections to the $W$ mass:  $\delta m_w = (2.9 \pm 1.6) \times 10^{-4}$.
We see that the  $Z$ and $W$ pole observables in Tables  \ref{tab:EWPT_zpole} and \ref{tab:EWPT_wpole}  simultaneously constrain all leptonic and heavy quark vertex corrections with (typically) per mille level accuracy. 
On the other hand, they {\em cannot} simultaneously constrain all light quark vertex corrections; in fact, only 3 linear combinations of  
$\delta g^{Zu}_L$, $\delta g^{Zu}_R$, $\delta g^{Zd}_L$ and $\delta g^{Zd}_R$ are probed by these observables. 
It is possible to show that the linear combination  
\beq
\label{eq:flat}
\delta g^{Zu}_L  + \delta g^{Zd}_L  +  {3 g_L^2 - g_Y^2 \over 4 g_Y^2} \delta g^{Zu}_R  
+ {3 g_L^2 + g_Y^2 \over 2 g_Y^2} \delta g^{Zd}_R 
\eeq 
is {\bf not} probed at all by the observables in Tables~\ref{tab:EWPT_zpole} and \ref{tab:EWPT_wpole}. 
In other words, it is a flat direction in the ${\cal O}(\Lambda^{-2})$ EFT fit. 
In order to characterize the constraints on the light quark couplings, it is convenient to introduce new variables $x,y,z,t$ related by a rotation to the light quark vertex corrections: 
\beq
\label{eq:xyztdefs}
\bvec x \\ y \\ z \\ t\evec = R  \bvec 
\delta g^{Zu}_L \\  \delta g^{Zu}_R \\  \delta g^{Zd}_L \\ \delta g^{Zd}_R  \evec
= 
\begin{pmatrix}
0.93 &  -0.29 &  -0.23 & -0.01  \\ 
0.18  &  0.87 &  -0.33 &  -0.33 \\ 
0.27  & 0.18  & 0.90 &  -0.29 \\ 
0.17  & 0.37  & 0.17  &  0.90 
\end{pmatrix} 
\bvec\delta g^{Zu}_L \\  \delta g^{Zu}_R \\  \delta g^{Zd}_L \\ \delta g^{Zd}_R  \evec 
, 
\eeq
where $R$ is an $SO(4)$ matrix such that: 1) $t$ is proportional to the flat direction in \eref{flat}; 
2) $x$, $y$, $z$ are mutually uncorrelated in the fit. 
The combinations $x,y,z$ are constrained by the above-discussed pole observables with percent-level accuracy: 
\beq
\label{eq:LEP1xyz}
\bvec 
x\\  y   \\   z 
\evec 
= \bvec  
-0.9 \pm 1.8 \\   
0.3  \pm 3.3 \\  
-2.4 \pm 4.8  
\evec \times 10^{-2} . 
\eeq 
As defined above, the fourth combination $t$ is not constrained by the observables in Tables~\ref{tab:EWPT_zpole} and \ref{tab:EWPT_wpole}.  
Another experimental input is needed to lift the flat direction, and thus to simultaneously constrain all four light quark vertex corrections. 
In Ref.~\cite{Efrati:2015eaa} the measurement of  $Zq\bar q$ couplings in Tevatron's D0~\cite{Abazov:2011ws} is employed to that purpose. 
However the resulting limit is very loose: 
$t = (0.1 \pm 17)\times 10^{-2}$, which is an order of magnitude weaker than the constraints in \eref{LEP1xyz} on the remaining three combinations.

In the following of this paper we show that leptonic $Z$ asymmetries at the LHC provide a more promising and more precise route to constrain $t$ and lift the flat direction in the SMEFT fit of $W$ and $Z$ pole observables.

\section{Hadron colliders as probes of $Zqq$ couplings} \vspace{0.4cm}
\label{sec:hadron}

The presence of the flat direction given in~\eref{flat} and the precise LHC/Tevatron extractions of the weak angle suggest that measurements of on-shell $Z$ production in hadron colliders can be important from the global SMEFT perspective, even if the accuracy is somewhat inferior compared to the most precise LEP-1 measurements. The goal of this section is to study this in detail.

\subsection{The Drell-Yan forward-backward asymmetry at the LHC} \vspace{0.4cm}

The natural LHC process to probe the $Z$ interactions with light quarks is the Drell-Yan production of lepton pairs, $p p  \rightarrow Z/\gamma^* \rightarrow l^+ l^-$. 
Indeed, at the parton level the intermediate $Z$ is produced via $q  \bar q \to Z$, providing tree-level sensitivity to the $Z q q$ couplings. 
We will focus on the FB asymmetry in the Drell-Yan production, which is a particularly clean observable thanks to cancellations of various QCD and PDF uncertainties, 
and for which high precision has been achieved both on the experimental and the theory side~\cite{Khachatryan:2016yte, Aaboud:2017ffb}. 
This observable has been used to extract with high accuracy the weak mixing angle~\cite{Sirunyan:2018swq}, and it is currently being discussed as a possible tool to further constrain the parton distribution functions~\cite{Accomando:2018nig,Bodek:2016olg}.

At the partonic level, the asymmetry appears in the decay angle $\theta^*$ of the negatively charged lepton with respect to the incoming quark direction in the center of mass frame. 
It is non-zero already in the SM, due to  parity-violating $Z$ couplings to fermions. In the SM the differential cross section of the partonic process $q \overline{q} \rightarrow Z/\gamma^* \rightarrow \ell^+ \ell^-$ is given at leading order by~\cite{Chatrchyan:2011ya}:
\bea
\label{eq:evenodd}
\frac{d\hat{\sigma}_{q \overline{q}}\left(\hat{s}, cos\theta^* \right) }{d\cos\theta^*}
& \propto& \hat{\sigma}^{even}_{q \overline{q}}\left(\hat{s}, \cos \theta^*  \right) + \hat{\sigma}^{odd}_{q \overline{q}}\left(\hat{s}, \cos \theta^*  \right) \nnl
& \propto& H^{even}_{q \overline{q}}\left(\hat{s} \right)\left(1+ \cos^2\theta^*  \right) + H^{odd}_{q \overline{q}}\left(\hat{s}\right)\cos\theta^*, 
\eea
where 
\bea
    H^{even}_{q \overline{q}} \left(\hat{s} \right)  &=& \frac{3}{2} \frac{\hat{s}}{\left(\hat{s}-m_{Z}^2 \right)^2 + m_Z^2\Gamma_Z^2} \left( \left( g^{Zq}_{V}\right)^2 + \left( g^{Zq}_{A}\right)^2 \right) \left( \left( g_V^{Zl}\right)^2 + \left( g_A^{Zl}\right)^2 \right) + 
    \nnl 
    && + \frac{3 \left( Q_q\right)^2\left( Q_l\right)^2}{2\hat{s}} + \frac{3\left(\hat{s}-m_{Z}^2 \right) }{\left(\hat{s}-m_{Z}^2 \right)^2 + m_Z^2\Gamma_Z^2} Q_q Q_l \, g^{Zq}_{V} g_V^{Zl},
    \\
    H^{odd}_{q \overline{q}} \left(\hat{s} \right) 
    &=&  \frac{12 \hat{s}}{\left(\hat{s}-m_{Z}^2 \right)^2 + m_Z^2\Gamma_Z^2} g^{Zq}_{V} g^{Zq}_{A} g_V^{Zl} g_A^{Zl}  + \frac{6\left(\hat{s}-m_{Z}^2 \right)}{\left(\hat{s}-m_{Z}^2 \right)^2 + m_Z^2\Gamma_Z^2}   Q_q Q_l \, g^{Zq}_{A} g_A^{Zl} .
\eea

Here,  $\hat{s}$ is the invariant mass of the dilepton system, $Q_f$ are the electric charges and $g_{V/A}^{Zf}\equiv g^{Zf}_R \pm g^{Zf}_L$ are the vector/axial couplings of the $Z$ boson to fermions. 
It is trivial to include the effect of SMEFT corrections to the $Zff$ couplings in the differential cross-sections given above. As long as we stay near the $Z$ pole ($\hat{s}\approx M_Z$) the corrections due to 4-fermion SMEFT operators are suppressed by $\Gamma_Z/m_Z \sim 0.02$, and can be neglected in the leading order approximation.\footnote{Note that finite (i.e. not log-enhanced) 1-loop SMEFT corrections, also neglected in our approximation, are suppressed by a numerically similar factor. 
}

The term proportional to $\cos \theta^*$ in~\eref{evenodd} will induce a difference in the number of events with a negative lepton going in the forward ($\cos \theta^*$ > 0) and backward ($\cos \theta^*$ < 0) directions.  
This asymmetry cannot be directly observed at the LHC because there we are not dealing with quarks as incoming particles, but with protons. Thus, we must modify~\eref{evenodd} by including parton distribution functions (PDFs). Moreover, having two identical particles in the initial state introduces an important complication: the absence of a preferred direction that one can use to build an asymmetry. In other words, we do not know {\it in every individual event} which  proton contains the quark and which one contains the antiquark that are annihilating. In order to circumvent this issue, the asymmetry is defined with respect to the longitudinal boost of the dilepton system on an event-by-event basis. Equivalently, one assumes that the momentum of the quark is larger than that of the antiquark~\cite{Chatrchyan:2011ya}, which is only true on average. This inevitably introduces a distorsion on the asymmetry at the parton level, which can be analytically parametrized through a dilution factor, as shown below.

An additional complication arises because (anti)quarks might have non-zero transverse momentum, which impedes us from equating the direction of the quark-antiquark pair in the partonic interaction to the beam direction. This effect can be minimized working in the so-called Collins-Soper frame~\cite{Collins:1977iv}, where $\theta^*$ is defined as the angle between $\ell^-$ and the axis that bisects the angle between the quark-momentum direction and the opposite directions to the antiquark momentum. 
This subtlety is not relevant for our work, which is restricted to studying the tree-level SMEFT corrections, in which case the transverse momentum of the incoming quarks is zero.

All in all, once these changes are implemented,~\eref{evenodd} transforms into~\cite{Chatrchyan:2011ya}:

\begin{equation} 
\label{eq:xsection}
\frac{d \sigma_{pp} \left(Y, \hat{s}, \cos\theta^* \right)}{dY\, d\hat{s} \, d \cos\theta^*} \propto \sum_{q=u,d,s,c,b}\!\!\! \left[ \hat{\sigma}^{even}_{q \overline{q}}\left(\hat{s}, \cos\theta^* \right) + D_{q \overline{q}}\left(Y, \hat{s} \right) \hat{\sigma}^{odd}_{q \overline{q}}\left(\hat{s}, \cos\theta^* \right) \right]F_{q \overline{q}}\left(Y, \hat{s} \right),
\end{equation}

where $Y$ is the rapidity of the dilepton center-of-mass system,
$F_{q \overline{q}}\left(Y,\hat{s} \right)$ is called the parton factor and $D_{q \overline{q}}\left(Y, \hat{s} \right)$ is the dilution factor to which we alluded previously. They depend on the PDFs as:

\begin{eqnarray}
F_{q \overline{q}} \left( Y,\hat{s} \right) 
&=& f_{q} \left( e^{+Y} \sqrt{\frac{\hat{s}}{s}}, \hat{s} \right) f_{\overline{q}} \left( e^{-Y} \sqrt{\frac{\hat{s}}{s}}, \hat{s} \right) + f_{q} \left( e^{-Y} \sqrt{\frac{\hat{s}}{s}}, \hat{s} \right) f_{\overline{q}} \left( e^{+Y} \sqrt{\frac{\hat{s}}{s}}, \hat{s} \right),
\\
D_{q \overline{q}} \left(Y, \hat{s} \right) 
&=& \frac{f_{q} \left( e^{+|Y|} \sqrt{\frac{\hat{s}}{s}}, \hat{s} \right) f_{\overline{q}} \left( e^{-|Y|} \sqrt{\frac{\hat{s}}{s}}, \hat{s} \right) - f_{q} \left( e^{-|Y|} \sqrt{\frac{\hat{s}}{s}}, \hat{s} \right) f_{\overline{q}} \left( e^{+|Y|} \sqrt{\frac{\hat{s}}{s}}, \hat{s} \right)}{F_{q \overline{q}}\left(\hat{s}, Y \right)} ,
\end{eqnarray}

\noindent where $s$ is the proton-proton invariant mass. Using this hadronic differential cross-section, the FB asymmetry is defined as:

\begin{equation}
\label{eq:AFBdef}
 A_{FB}\left( Y, \hat{s}\right) = \frac{\sigma_F \left( Y, \hat{s}\right)-\sigma_B \left( Y, \hat{s}\right)}{\sigma_F \left( Y, \hat{s}\right)+\sigma_B \left( Y, \hat{s}\right)},  
\end{equation}

\noindent where the forward and backward cross-sections, $\sigma_F$ and $\sigma_B$, are obtained by integrating the differential cross section over the positive and negative values of $\cos\theta^*$, respectively. It should also be noted that, to calculate  $A_{FB}$ integrated  over the $Y$ and $\hat{s}$ bins, one should integrate independently $\sigma_F$ and $\sigma_B$ and then calculate the integrated $A_{FB}$ from that input. 

The dependence of the asymmetry on the invariant mass and rapidity of the dilepton system effectively increase the number of independent observables at our disposal.
In this work we restrict to the dilepton masses close to the $Z$ peak, so that contributions from 4-fermion operators can be neglected and only vertex corrections need to be considered. Corrections to the leptonic $Z$ couplings can also be neglected, due to the very stringent LEP-1 constraints shown in~\eref{dgZl}.
The effects due to the vertex correction involving the  $s$, $b$ and $c$  are suppressed by the small PDFs of the heavy quarks in the proton, and again can be neglected given the LEP-1 constraints.  
Using measurements of the asymmetry at different rapidity bins, we will be able to probe different combinations of $Zqq$ couplings. 
In principle, four distinct rapidity bins are enough to disentangle all four $\delta g_{L/R}^{Zd/Zu}$ corrections, although this may be hindered in practice by large correlations between the bins, as we will see in the following.

\subsection{Numerical analysis} \vspace{0.4cm}

In this section we set bounds on the $Z$ couplings to light quarks using the measurement of the ATLAS collaboration of the angular distributions of leptons from the Drell-Yan process in the $\sqrt{s}=8$ TeV data~\cite{ATLAS:2018gqq}. In particular we will use  the measurement of the so-called $A_4$ angular coefficient, which is defined upon expanding the Drell-Yan differential cross section in harmonic polynomials, and it is related to the FB asymmetry by  $A_{FB} = (3/8) A_4$. This equation holds at all orders in QCD when considering the full phase space of the individual decay leptons. Thus, the results of the previous sections are immediately applicable to this measurement, up to the trivial $3/8$ numerical factor. 

Table~\ref{tab:A4meas} shows the results of the ATLAS measurements of $A_4$ for dileptons at the $Z$ peak ($80~\rm{ GeV} < \sqrt{\hat{s}} < 100~\rm{ GeV}$). As we can see, the asymmetry was measured for four values of the dilepton rapidity, which will allow us to separate (at least formally) the four corrections of the $Z$ boson to up and down quarks. 
Table~\ref{tab:A4meas} also shows the SM predictions, as given in the ATLAS paper~\cite{ATLAS:2018gqq}, which were calculated at NNLO in QCD~\cite{Bozzi:2010xn, Catani:2015vma, Catani:2009sm}. 
We can see that, contrary to the electroweak observables discussed in~\sref{exp}, the theory errors are comparable to the  experimental ones and cannot be neglected.

\begin{table}
 \begin{center}
 \begin{tabular}{|c|c|c|}
 \hline
{|Y|} & {Experimental value}   
&  {SM prediction}    
 \\  \hline 
0.0 - 0.8  & $ 0.0195 \pm 0.0015 $    &  $ 0.0144 \pm 0.0007$   
\\ \hline 
0.8 - 1.6  & $ 0.0448 \pm 0.0016 $    &  $ 0.0471 \pm 0.0017$   
\\ \hline 
1.6 - 2.5  & $ 0.0923 \pm 0.0026 $    &  $ 0.0928 \pm 0.0021$   
\\ \hline 
2.5 - 3.6  & $ 0.1445 \pm 0.0046 $    &  $ 0.1464 \pm 0.0021$   
\\ \hline 
\end{tabular}
\end{center}
\caption{ATLAS $A_4$ measurements obtained in Ref.~\cite{ATLAS:2018gqq} using $\sqrt{s}=8$ TeV data for a dilepton invariant mass in the range ($80~\rm{ GeV}, 100~\rm{ GeV}$). The SM predictions, obtained at NNLO in QCD, are also taken from Ref.~\cite{ATLAS:2018gqq}. 
}
\label{tab:A4meas}
 \end{table}

The  theory predictions are modified in the presence of nonstandard corrections. Working at linear order in them we have:
\bea
\label{eq:Ath}
A^{th}_{4,i} = A^{SM}_{4,i}\left(1  + \alpha_{ik} \,\delta g_k \right)
\eea
where the $i$ index corresponds to the four rapidity bins (see~\tref{A4meas}) and $\delta g_k = \{ g^{Zu}_L, g^{Zu}_R,$ $g^{Zd}_L, g^{Zd}_R\}$. 
The numerical factors $\alpha_{ik}$ are calculated using Eqs.~\eqref{eq:xsection}-\eqref{eq:AFBdef}, after integration over the $\hat{s}$ and $Y$ bins and using the MMHT2014lo68cl PDF set~\cite{Harland-Lang:2014zoa}.
The uncertainties of this PDF set and those obtained changing the PDF set were calculated but, since their impact on our final results is negligible, we will ignore them hereafter.

We use this semi-analytical approach, 
instead of a purely numerical one, because the cross section we are interested in is free of kinematic cuts, except for those on the dilepton invariant mass and rapidity. This follows from the definition of the $A_4$ angular coefficient, which includes an integration over the whole individual lepton phase space, which is carried out experimentally through an unfolding procedure~\cite{ATLAS:2018gqq}. Our semi-analytical approach simplifies the analysis and avoids uncertainties from the numerical simulation, which may be significant due to the small size of the FB asymmetry. Nonetheless, we have crosschecked our results using a numerical approach, in which the cross sections are calculated using MadGraph~\cite{madgraph:2014alwall} and processed by MadAnalysis5~\cite{madanalysis1}. Neglected contributions from parton shower or detector effects are expected to be small for this process, which only involves light leptons in the final state. 

With the results in~\tref{A4meas} and~\eref{Ath} we build the following $\chi^2$ function:
\begin{equation}
\label{eq:chi2}
\chi^2 \left( \delta g_k \right) = \sum_{i} \left(\frac{A^{SM}_{4,i}\left(1  + \alpha_{ik} \delta g_k \right) - A^{exp}_{4,i}}{\delta A_{4,i}} \right)^2,
\end{equation}
\noindent 
where the sum runs over the four bins of rapidity displayed in \tref{A4meas}. Experimental and SM errors are combined in quadrature in $\delta A_{4,i}$.

Before minimizing the full $\chi^2$ function, it is instructive to see how each rapidity bin of the $A_{4}$ measurement  performs in restricting the EFT couplings. 
We find:

\bea 
\label{eq:binbybin}
0.0<|Y|<0.8:~&& 0.63\, \delta g^{Zu}_{L} + 0.71 \,\delta g^{Zu}_{R} - 0.20 \,\delta g^{Zd}_{L} - 0.22 \,\delta g^{Zd}_{R} = 0.088(29)~,
\nnl 
0.8<|Y|<1.6:~&& 0.60 \,\delta g^{Zu}_{L} + 0.74 \,\delta g^{Zu}_{R} - 0.18 \,\delta g^{Zd}_{L} - 0.22 \,\delta g^{Zd}_{R} = -0.012(12)~,
\nnl 
1.6<|Y|<2.5:~&&  0.53 \,\delta g^{Zu}_{L} + 0.80 \,\delta g^{Zu}_{R} - 0.16 \,\delta g^{Zd}_{L} - 0.23 \,\delta g^{Zd}_{R} = -0.0014(92)~,
\nnl 
2.5<|Y|<3.6:~&&  0.43 \,\delta g^{Zu}_{L} + 0.86 \,\delta g^{Zu}_{R} - 0.18 \,\delta g^{Zd}_{L} - 0.21 \,\delta g^{Zd}_{R} = -0.0030(81)~,~~~~~ 
\eea
\noindent where we have normalized each combination to ease the comparison. 
We can see that the limits reach a percent-level accuracy.
The limits from the lowest rapidity bin are the weakest because it is affected the most by the smearing out of the asymmetry: in this bin the direction of the boost of the dilepton system is often opposite to the direction of the incoming quark.

Let us now minimize the full likelihood in~\eref{chi2}, i.e., we combine the 
four constraints in \eref{binbybin}. 
We can see by eye that each bin constrains a similar linear combination of the quark couplings.
Therefore we can already anticipate that attempts to constrain simultaneously all four vertex corrections using just the ATLAS data will suffer from large correlations. 
Instead, it is more instructive to show the constraints on the four linear combinations that are orthonormal and uncorrelated, which we denote as $\{ x',y',z',t'\}$, similarly to our approach in~\sref{exp}. We find:

\begin{equation}
\label{eq:AFBprimes}
\left( \begin{matrix}
x' = 0.21 \delta g_{L}^{Zu} + 0.19 \delta g_{R}^{Zu} + 0.46 \delta g_{L}^{Zd} + 0.84 \delta g_{R}^{Zd}
\\ 
y' = 0.03 \delta g_{L}^{Zu} - 0.07 \delta g_{R}^{Zu} - 0.87 \delta g_{L}^{Zd} + 0.49 \delta g_{R}^{Zd}
\\ 
z' = 0.83 \delta g_{L}^{Zu} - 0.54 \delta g_{R}^{Zu} + 0.02 \delta g_{L}^{Zd} - 0.10 \delta g_{R}^{Zd}
\\
t' = 0.51 \delta g_{L}^{Zu} + 0.82 \delta g_{R}^{Zu} - 0.17 \delta g_{L}^{Zd} - 0.22 \delta g_{R}^{Zd}
\end{matrix} \right) = \left( \begin{matrix}
-10 \pm 4 \\ 
0.5 \pm 0.4  \\ 
0.04 \pm 0.06 \\ 
-0.001 \pm 0.005  
\end{matrix} \right)~.
\end{equation}

As anticipated, only a single direction, which we denote as $t'$, is constrained at $0.5\%$ level. 
The constraints on the directions orthogonal to $t'$ are much weaker,
and those on $x'$ are practically void within the EFT validity regime.
This means that the ATLAS data alone cannot give us useful constraints on all $Z$ boson couplings to quarks.
Nevertheless, the input from ATLAS will be invaluable once combined with the other probes of the $Z q q$ couplings, 
as we will quantify in the next subsection. 

\subsection{Impact of $A_4$ on the global fit} \vspace{0.4cm}

Combining the $A_{FB}$ limits obtained above with those extracted in~\sref{exp} from traditional pole data we find:
\begin{equation}
\label{eq:LEPplusLHC}
\left( \begin{matrix}x \\ y \\ z \\t \end{matrix} \right) = \left( \begin{matrix}0.004 \pm 0.017 \\ 0.010 \pm 0.032  \\ 0.021 \pm 0.046 \\ -0.03 \pm 0.19  \end{matrix} \right), \quad 
\rho = 
\left(
\begin{array}{cccc}
 1. & -0.09 & -0.08 & -0.04 \\
 -0.09 & 1. & -0.09 & -0.93 \\
 -0.08 & -0.09 & 1. & -0.19 \\
 -0.04 & -0.93 & -0.19 & 1. \\
\end{array}
\right).
\end{equation}

\begin{figure}
\centering
\includegraphics[scale=0.475]{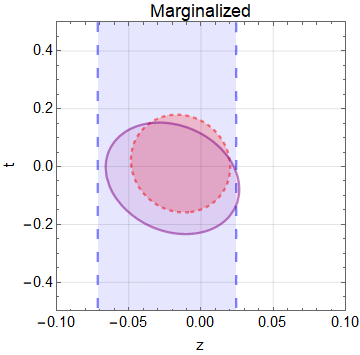}
\includegraphics[scale=0.455]{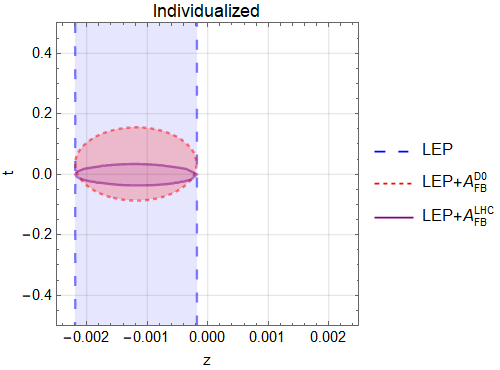} 

\caption{Allowed regions (corresponding to $\Delta\chi^2=1$) for the $z$ and $t$ combinations of vertex corrections, see Eqs.~\eqref{eq:xyztdefs} for their definition. The blue vertical band is obtained using LEP data and the rest of traditional pole observables discussed in~\sref{exp}, whereas the purple and orange ellipses add the FB asymmetry measurements at LHC and D0 respectively (see main text for details). The left panel is obtained marginalizing over the remaining EFT parameters, {\it cf.}~\eref{TH_dg}, whereas the right panel is obtained setting all of them to zero. Note the horizontal scale is not the same in both plots. } 
\label{fig:ztplot}
\end{figure}

Comparing with~\eref{LEP1xyz}, we see that the marginalized bounds on $x$, $y$ and $z$ are essentially the same, but the $t$ combination is not anymore unconstrained. The per mille level $A_{FB}$ constraint on $t'$, {\it cf.}~\eref{AFBprimes}, generates a much weaker bound on $t$ because these combinations of $Z$ couplings happen to be quite orthogonal ($t\cdot t' = 0.16$). In spite of this, the $A_{FB}$ bound on $t$ is stringent enough to make a significant impact when added to the traditional pole data. In~\eref{LEPplusLHC} this is reflected by the large correlation between $y$ and $t$, which was zero before the inclusion of the $A_{FB}$ data. This is illustrated in Fig.~\ref{fig:ztplot}, which shows the impact of the inclusion of the FB asymmetry on the $z$ and $t$ combinations of vertex corrections.  

Beyond the numerical impact on the extracted bounds, the inclusion of the FB asymmetry in the analysis strengthens the robustness of the EFT approach. In fact, the inclusion of $(\delta g)^2$ corrections in the analysis would have a major impact on both LEP and LHC results since ${\cal O}(1)$ corrections cannot be excluded by each dataset separately. However, such large corrections are not anymore possible once both datasets are used together. This is clearly illustrated in Fig.~\ref{fig:ztplot}.

\begin{figure}[b]
\centering
\includegraphics[scale=0.34]{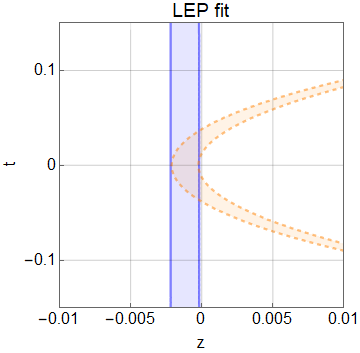} 
\includegraphics[scale=0.34]{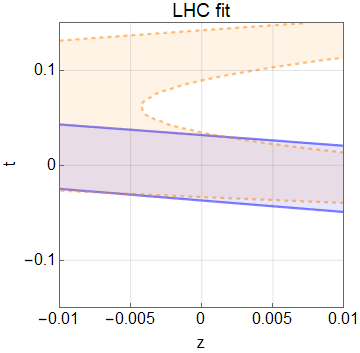} 
\includegraphics[scale=0.34]{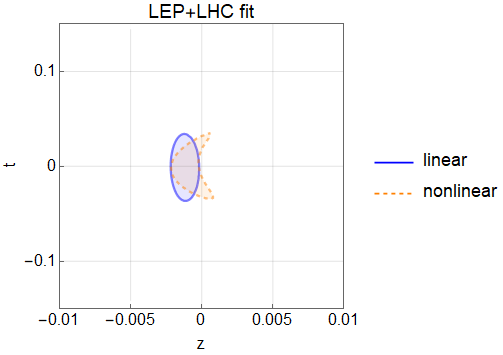} 
\caption{Comparison between linear and nonlinear contours ($\Delta\chi^2=1$) in a 2D fit using the ``traditional" pole observables discussed in~\sref{exp} (LEP fit, left panel), the ATLAS FB asymmetry data (LHC fit, central panel) and the combination of both (right panel).} 
\label{fig:ztplotlinearvsnonlinear}
\end{figure}

The importance of $A_{FB}$ data is even clearer in simple New Physics scenarios where only some $Zqq$ couplings are modified. This was anticipated in~\sref{intro} through Fig.~\ref{fig:deltag}, which shows the constraints on the modifications of the $Z$ couplings to up quarks ($\delta g_L^{Zu}$ and $\delta g_R^{Zu}$) assuming all other nonstandard contributions can be neglected. 
\fref{deltagallpairs} shows that the situation is similar for other pairs of vertex corrections. The fact that $A^{LHC}_{FB}$ provides crucial information in such simple scenarios despite the extremely precise LEP measurements is highly nontrivial and represents one of our main findings. 

\begin{figure}[h]
\centering \hspace{-2.2cm}
\includegraphics[scale=0.493]{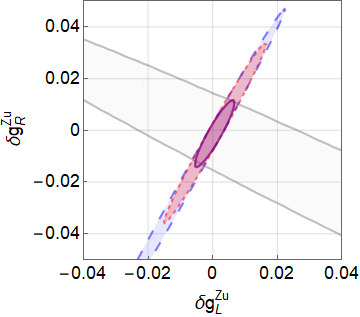}
\includegraphics[scale=0.48]{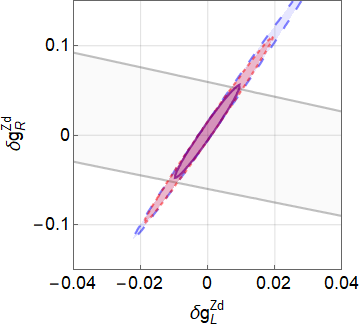} \\
\includegraphics[scale=0.47]{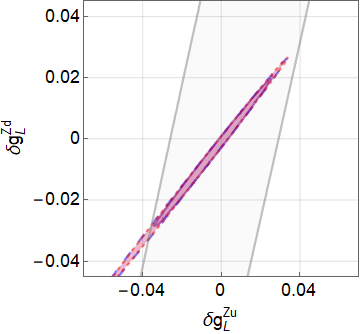} \hspace{0.1cm}
\includegraphics[scale=0.452]{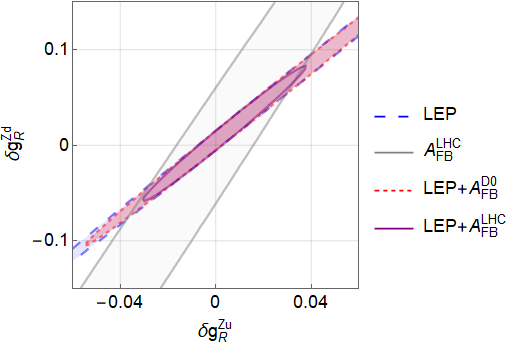}
\caption{Allowed regions (at 95\% CL) for four pairs of corrections of the $Z$ couplings to light quarks. In each case, only that pair of corrections is different from zero, the rest are given by their SM values. Note the scale is different in each panel.}
\label{fig:deltagallpairs}
\end{figure}

\subsection{D$0$ measurement} \vspace{0.4cm}

As mentioned  in~\sref{exp}, Ref.~\cite{Efrati:2015eaa} included in their dataset the determination of the light quark couplings to the $Z$ boson at D0~\cite{Abazov:2011ws}. This allowed them to lift the flat direction that appears when only ``traditional" pole observables are used. Such determination of the light quark couplings was obtained by D0 from the FB asymmetry in $p\bar{p}\to e^+e^-$ with 5 fb$^{-1}$ at $\sqrt{s}=1.96$ TeV~\cite{Abazov:2011ws}. That extraction used dilepton masses up to $1$ TeV, so technically speaking it assumes that 4-fermion operators are absent or much smaller than vertex corrections. 
However, one expects that the D0 determination is dominated by the bins near the $Z$-pole, which have the largest statistics. This is confirmed by our simulations. For this reason, the removal of off-peak bins is not expected to change the D0 extraction of the vertex corrections.

Such D0 results are sensitive to individual vertex corrections at the $2$\%-level or worse, which is $\sim 4$ times weaker than the sensitivity of the LHC $A_{FB}$ data, {\it cf.}~\eref{AFBprimes}.
Combining the traditional observables from \sref{exp} with the D0 input we obtain:

\begin{equation}
\label{eq:D0plusLEP}
\left( \begin{matrix}x \\ y \\ z \\t \end{matrix} \right) = \left( \begin{matrix} -0.009 \pm 0.017 \\ 0.007 \pm 0.023  \\ -0.014 \pm 0.034 \\ 0.01 \pm 0.17  \end{matrix} \right), \quad 
\rho 
= \left( \begin{matrix}1 & 
-0.19 & 0.18 & 0.00 \\ 
-0.19 & 1 & 0.04 & -0.68 \\ 
0.18 & 0.04 & 1 & -0.07 \\ 
0.00 &-0.68 & -0.07 & 1 \end{matrix} \right).
\end{equation}

The comparison of the D0 limits with the LHC+LEP results of~\eref{LEPplusLHC} is complicated because one has to compare marginalized bounds {\it and} correlations. The former are more stringent in the D0+LEP case, but the latter are higher in the LHC+LEP case, which reflects the fact that the LHC  constrains a specific combination much strongly than D0. This means that both LHC and D0 measurements of the $A_{FB}$ asymmetries bring relevant information to the global fit, although for simple scenarios the LHC will typically have a more important effect, as shown in~\fref{deltag},~\fref{ztplot} (right panel) and~\fref{deltagallpairs}.

However, several caveats should be made concerning the D0 extractions. First, the limits will slightly weaken if one only uses the $A_{FB}$ measurement at the $Z$ pole. Since D0 does not study the dependence of the FB asymmetry with the dilepton rapidity, there is only one measurement at the $Z$ pole. Thus, one will be able to probe only one combination of couplings, contrary to the ATLAS case. Additionally, LHC collaborations are still active and taking data, so one can expect significant improvements on that front.

\subsection{Combined fit results} \vspace{0.4cm}

We close this section with a combined fit that includes ``traditional" pole observables presented in~\sref{exp} (mainly LEP) as well as the above-discussed measurements of the FB asymmetries by ATLAS and D0. The values obtained for the $(x,y,z,t)$ variables are the following:

\begin{equation}
\label{eq:D0plusLEPplusLHC}
\left( \begin{matrix}x \\ y \\ z \\t \end{matrix} \right) = \left( \begin{matrix}-0.005 \pm 0.016 \\ 0.009 \pm 0.022  \\ -0.014 \pm 0.032 \\ -0.03 \pm 0.13  \end{matrix} \right), \quad 
\rho 
= \left(
\begin{array}{cccc}
 1. & -0.25 & 0.1 & 0.01 \\
 -0.25 & 1. & -0.03 & -0.91 \\
 0.1 & -0.03 & 1. & -0.26 \\
 0.01 & -0.91 & -0.26 & 1. \\
\end{array}
\right), 
\end{equation}
Using \eref{xyztdefs} we translate this into constraints on the corrections of the $Z$ couplings to light quarks, which after all are the main subject of our study:
\begin{equation}
\label{eq:D0plusLEPplusLHC2}
\left( \begin{matrix}\delta g^{Zu}_{L} \\ \delta g^{Zu}_{R} \\ \delta g^{Zd}_{L} \\\delta g^{Zd}_{R} \end{matrix} \right) = \left( \begin{matrix}-0.012 \pm 0.024 \\ -0.005 \pm 0.032  \\ -0.020 \pm 0.037 \\ -0.03 \pm 0.13  \end{matrix} \right), \quad 
\rho 
= \left( \begin{matrix}1 & 
0.51 & 0.68 & 0.69 \\ 
0.51 & 1 & 0.56 & 0.94 \\ 
0.68 & 0.56 & 1 & 0.54 \\ 
0.69 & 0.94 & 0.54 & 1 \end{matrix} \right).
\end{equation}

These values are marginalized over the 16 remaining EFT parameters that also contribute to the fit observables, {\it cf.}~\eref{TH_dg}. The full set of constraints and the $20x20$ correlation matrix is given in~\aref{fullfit}.

\section{Conclusions and Discussion} \vspace{0.4cm}
\label{sec:conclusions}

In this  work we have discussed the impact of LHC $Z$-pole measurements on constraining the Wilson coefficients of dimension-6 operators in the SMEFT (mainly vertex corrections). 
Naively, in this domain the LHC should not be able to compete with LEP because the latter measured multiple $Z$-pole observables with per mille precision in a cleaner environment. 
However, LEP leaves one unconstrained direction among $Z$ boson couplings to up and down quarks.  
Indeed, one of the linear combinations defined in \eref{xyztdefs} and denoted as $t$ is not probed at all by LEP observables and represents a flat direction in the ${\cal O}(\Lambda^{-2})$ electroweak fit.   
Information from hadron colliders allows one to plug this hole. 
In particular, the Drell-Yan $pp \to \ell^+ \ell^-$ and $p \bar p \to \ell^+ \ell^-$  processes are sensitive to the $Z$ boson coupling to light quarks making up the proton and anti-proton. 
Most importantly for our sake, they are sensitive to a {\em different} linear combination of these couplings. 
Therefore, Drell-Yan measurements in hadron colliders are complementary to the LEP  observables and provide precious inputs to the electroweak fit. 
In this paper we exemplified this general fact using the FB asymmetry in $pp \to Z \gamma^* \to \ell^+ \ell^-$  measured by the ATLAS experiment.
One of our main results is that the flat direction along the $t$ variable is indeed lifted with the inclusion of the ATLAS input, {\it cf.}~\eref{LEPplusLHC}. Interestingly enough, we find that the ATLAS $A_{FB}$ information provides a significant improvement on LEP-only bounds on the $Zqq$ vertex corrections even in simple scenarios with few free parameters, as shown in~\fref{deltag}.

\begin{figure}[t]
\centering
\includegraphics[scale=0.7]{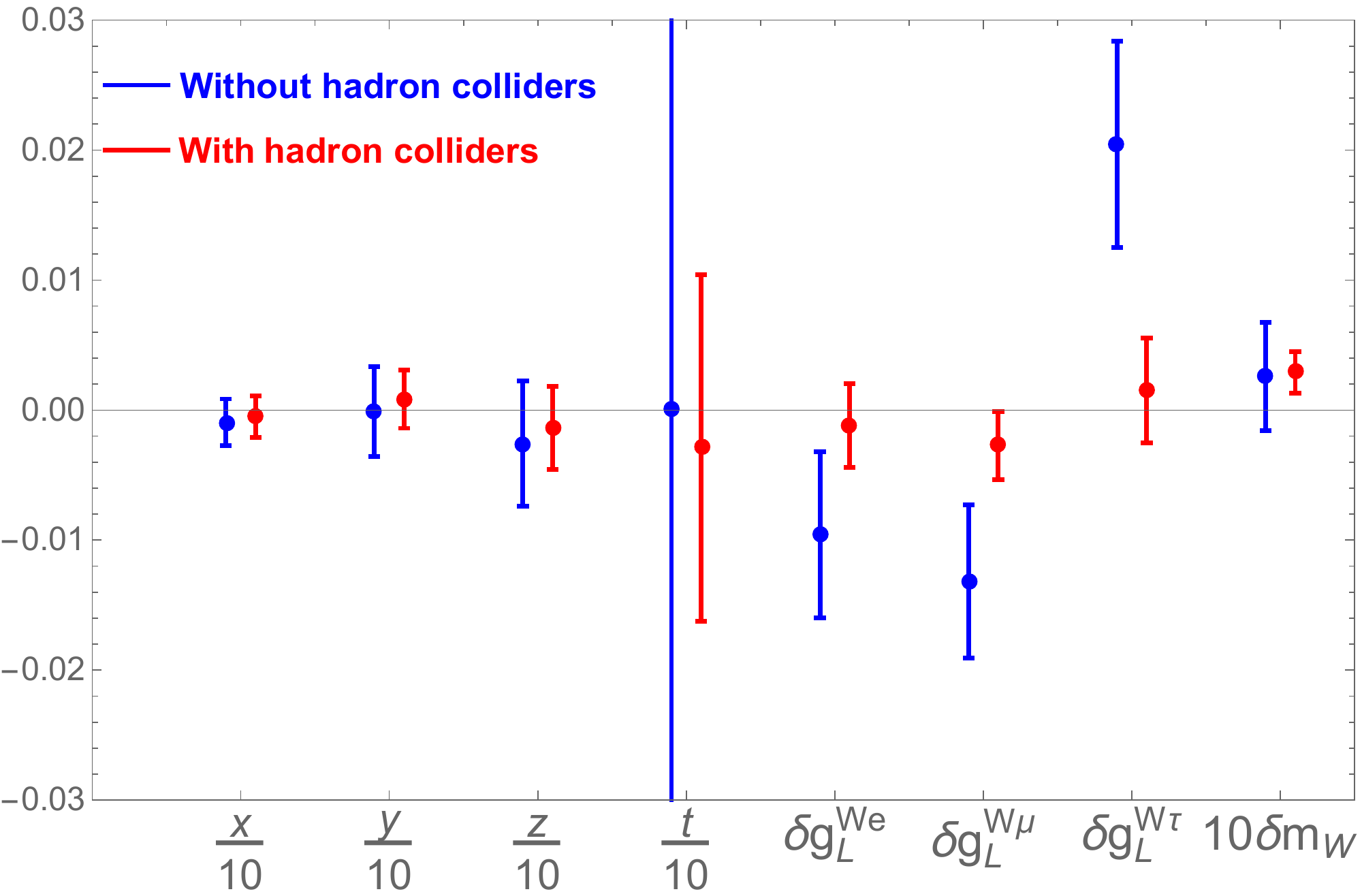}
\caption{
Comparison of the marginalized constraints on selected parameters in the electroweak fit, with (red error bars) and without (blue error bars) using hadron collider observables. 
Here, $x,y,z,t$  defined in \eref{xyztdefs} are linear combinations of the non-SM corrections to the coupling between the $Z$ boson and light quarks, 
$\delta g^{W\ell}_L$ are corrections to the coupling between the $W$ boson and   left-handed leptons, and $\delta m_W$ is a correction to the $W$ boson mass.
}
\label{fig:withorwithotyou}
\end{figure}

We would like to remark that the importance of hadron colliders for the electroweak fit is more general and goes beyond improving our knowledge of the $Zuu$ and $Zdd$ couplings.
This is illustrated in~\fref{withorwithotyou}, where we compare constraints on selected parameters with and without the use of hadron collider observables.
First, as shown in this work, constraints  on other combinations of $Z q \bar q$ couplings, which are denoted as $x,y,z$ and which are probed by LEP, are significantly improved by including information from FB asymmetry in lepton Drell-Yan production. 
Even more spectacular improvements happen in the $W$ boson sector. 
Indeed, LEP could produce $W^\pm$ only in the LEP-2 phase and in much lower numbers than $Z$, resulting in a lower accuracy of the $W$ observables.  
It is well known the precision with which the $W$ mass was measured at LEP-2 has been surpassed by the Tevatron and the LHC, which manage to squeeze down the error bar by almost a factor of three. 
It may be less known that a similar improvement has been achieved for the $W$ boson couplings to leptons. 
Thanks mainly to the recent measurements of leptonic branching fractions of $W$ in  ATLAS and LHCb~\cite{Aaij:2016qqz,Aaboud:2016btc,Aad:2020ayz} the error bars on the parameters $\delta g^{W\ell}_L$ are improved by approximately a factor of two for $\ell = e,\mu,\tau$. 
Moreover, tensions  with the SM visible in the LEP measurements, which in the past were subject to some theoretical scrutiny~\cite{Dermisek:2008dq,Filipuzzi:2012mg,Mader:2012pm}, are now  all gone away. 

We close with a comment on the perspectives of improving the electroweak fit using future results.  
Currently, the error for the ATLAS $A_4$ measurement (used in our study) is dominated by statistics, therefore the bounds on $Z qq$ couplings should be tightened once more data is analyzed. 
The current and future measurements of Drell-Yan dilepton  production by CMS and LHCb could be analyzed following a similar procedure to ours. 
It is worth noting that Drell-Yan production at the Tevatron, which was a $p \bar p$ collider, probes a different combination of the $Z qq$ vertex correction, therefore it would be advantageous to include the final 10~fb$^{-1}$ Tevatron results in the fit as well.  
Ideally, all these analyses would be done by the experimental collaboration themselves, who are in the best position to assess the pertinent systematic errors and correlations. 
Next, information from Drell-Yan cross sections (in addition  to asymmetries) could be added, and the off-pole  data could be analyzed at the same time in the context of a more general fit to both vertex corrections and 4-fermion operators~\cite{Falkowski:2017pss,Horne:2020pot}. 
Our constraints do not include vertex corrections to the $Ztt$ coupling, but these can be accommodated too in a more comprehensive fit~\cite{Buckley:2015nca,AguilarSaavedra:2018nen,Ellis:2020unq,Bruggisser:2021duo}. 
Finally, it would be interesting to analyze the potential of the High-Luminosity LHC~\cite{ATLAS:2018qvs} and other future colliders for improving the constraints on $Zqq$ vertex corrections.  
We expect that these future studies will increase even further the impact of hadron colliders on the electroweak precision program.

\begin{acknowledgments}
 VB is supported by by Ministerio de Ciencia, Innovación y Universidades, Spain [grant FPU18/01340]. AF has received funding from the Agence Nationale de la Recherche (ANR) under grant ANR-19-CE31-0012 (project MORA) and from the European Union’s Horizon 2020 research and innovation programme under the Marie Skłodowska-Curie grant agreement No 860881-HIDDeN. 
MGA is supported by the {\it Generalitat Valenciana} (Spain) through the {\it plan GenT} program (CIDEGENT/2018/014).

\end{acknowledgments}

\newpage
\appendix

\section{Map to the Warsaw basis} \vspace{0.4cm}
\label{app:warsaw}

The results of this paper were presented as constraints on the subset of independent vertex corrections defined by the Lagrangian in \eref{TH_Lvff}. 
The vertex corrections can be expressed as linear combinations of the Wilson coefficients in any basis of the SMEFT.
In this appendix we write down the map to the Warsaw basis.\footnote{For the map to the SILH basis~\cite{Contino:2013kra} see Ref.~\cite{deFlorian:2016spz}.}   
Adopting the conventions and notation of {\em Wilson coefficient exchange format} (WCxf)~\cite{Aebischer:2017ugx}, the relationship between the independent vertex corrections and the Warsaw basis Wilson coefficients is given by
\bea
\label{eq:DEF_dgMap}
v^{-2}  \delta g^{W e}_L & = &   C^{(3)}_{\varphi l} + f(1/2,0) - f(-1/2,-1),  
\nnl 
v^{-2} \delta g^{Ze}_L & = &    - {1 \over 2} C^{(3)}_{\varphi l} - {1\over 2} C_{\varphi l}^{(1)}+   f(-1/2, -1) , 
\nnl 
v^{-2}  \delta g^{Ze}_R & = &  - {1\over 2} C_{\varphi e}^{(1)}   +  f(0, -1) ,
\nnl 
v^{-2}  \delta g^{Z u}_L & = &  {1 \over 2} V C^{(3)}_{\varphi q} V^\dagger - {1\over 2} V C_{\varphi q}^{(1)}V^\dagger   + f(1/2,2/3) ,  
\nnl
v^{-2}  \delta g^{Zd}_L & = &   
 - {1 \over 2}  C^{(3)}_{\varphi q} - {1\over 2} C_{\varphi q}^{(1)}   + f(-1/2,-1/3), 
\nnl
v^{-2}  \delta g^{Zu}_R & = &    - {1\over 2} C_{\varphi u}   +  f(0,2/3), 
\nnl
v^{-2}  \delta g^{Zd}_R & = &    - {1\over 2} C_{\varphi d}  +  f(0,-1/3), 
\eea 
where 
\beq 
\label{eq:DEF_fdeltag}
f(T^3,Q)  \equiv  \bigg \{  
-   Q  {g_L  g_Y \over  g_L^2 -  g_Y^2} C_{\varphi WB} 
 -  \left ( 
{1 \over 4} C_{\varphi D}  +  {1 \over 2 } \Delta_{G_F}  \right )  \left ( T^3 + Q { g_Y^2 \over  g_L^2 -   g_Y^2} \right )  \bigg \} {\bf 1} ,  
\eeq
and $\Delta_{G_F} \equiv [C^{(3)}_{\varphi l}]_{11} + [C^{(3)}_{\varphi l}]_{22} -  {1 \over 2}[C_{ll}]_{1221}$.
Moreover, $\delta m_w =  {1 \over 2} [\delta g^{We}_L]_{11} + {1 \over 2} [\delta g^{We}_L]_{22}  - {v^2 \over 4} [C_{ll}]_{1221}$.

\section{Complete results and correlation matrix} 
\label{app:fullfit}
The marginalized constraints on all 20 independent parameters in our fit read
\beq 
\label{eq:resultsFinal}
\bvec 
\delta g^{We}_L \\ \delta g^{W\mu}_L \\ \delta g^{W\tau}_L \\ 
\delta g^{Ze}_L \\ \delta g^{Z\mu}_L \\ \delta g^{Z\tau}_L \\ 
\delta g^{Ze}_R \\ \delta g^{Z\mu}_R \\ \delta g^{Z\tau}_R \\ 
\delta g^{Zu}_L \\ \delta g^{Zu}_R \\ 
\delta g^{Zd}_L \\ \delta g^{Zd}_R \\ 
\delta g^{Zs}_L \\ \delta g^{Zs}_R \\ 
\delta g^{Zc}_L \\ \delta g^{Zc}_R \\ 
\delta g^{Zb}_L \\ \delta g^{Zb}_R \\ 
\delta m_w 
\evec  \qquad =  \qquad \bvec 
-1.2\pm 3.2 \\ -2.7\pm 2.6 \\ 1.5\pm 4.0 \\ 
-0.20\pm 0.28\\ 0.1\pm 1.2 \\ -0.09\pm 0.59 \\ 
-0.43 \pm 0.27 \\ 0.0 \pm 1.4 \\ 0.62 \pm 0.62 \\ 
-12 \pm 23 \\ -4 \pm 31 \\ 
-19 \pm 36 \\ -30\pm 130  \\ 
11\pm 28  \\ 32 \pm 48 \\ 
-1.5\pm 3.6 \\ -3.3\pm 5.3 \\ 
3.1\pm 1.7 \\ 21.9\pm  8.8 \\
0.29 \pm 0.16 \evec  \times 10^{-3}. 
\eeq 
The correlation matrix is 
\begin{tiny}
\beq
\label{eq:correlationFinal}
\left(
\begin{array}{cccccccccccccccccccc}
 1 & 0.2 & -0.59 & -0.22 & -0.09 & 0.01 & 0.16 & -0.13 & -0.08 & -0.04 & -0.06 & -0.03 & -0.06 & -0.02 & -0.04 & -0.01 & 0.01 & 0.04 & 0.01 & 0. \\
 - & 1 & -0.39 & -0.27 & -0.11 & 0.01 & 0.2 & -0.16 & -0.1 & -0.04 & -0.06 & -0.03 & -0.07 & -0.03 & -0.04 & -0.01 & 0.01 & 0.05 & 0.01 & 0. \\
 - & - & 1 & -0.18 & -0.07 & 0.01 & 0.13 & -0.11 & -0.07 & 0. & 0. & 0. & 0. & 0. & 0. & -0.01 & 0. & 0.04 & 0.01 & 0. \\
 - & - & - & 1 & -0.09 & -0.07 & 0.16 & -0.04 & 0.04 & 0.04 & 0.06 & 0.03 & 0.06 & 0.03 & 0.04 & 0.07 & 0.08 & -0.36 & -0.35 & 0. \\
 - & - & - & - & 1 & 0.06 & -0.04 & 0.91 & -0.04 & 0. & 0.01 & 0.01 & 0.01 & 0.01 & 0.01 & -0.02 & -0.01 & 0.06 & 0.04 & 0. \\
 - & - & - & - & - & 1 & 0.02 & -0.03 & 0.41 & -0.01 & -0.01 & -0.01 & -0.02 & 0. & -0.01 & -0.02 & 0.01 & 0.07 & 0.01 & 0. \\
 - & - & - & - & - & - & 1 & -0.07 & -0.04 & -0.02 & -0.03 & -0.02 & -0.03 & -0.01 & -0.02 & 0.06 & 0.11 & -0.34 & -0.38 & 0. \\
 - & - & - & - & - & - & - & 1 & 0.04 & 0.02 & 0.03 & 0.02 & 0.03 & 0.01 & 0.02 & 0. & -0.01 & 0.01 & 0.03 & 0. \\
  - & - & - & - & - & - & - & - & 1 & 0.02 & 0.03 & 0.01 & 0.03 & 0.01 & 0.02 & 0.01 & -0.01 & -0.04 & 0. & 0. \\
 - & - & - & - & - & - & - & - & - & 1 & 0.5 & 0.68 & 0.69 & 0.07 & -0.29 & -0.05 & 0.09 & -0.02 & -0.01 & 0. \\
 - & - & - & - & - & - & - & - & - & - & 1 & 0.55 & 0.94 & -0.11 & -0.39 & 0.07 & 0.07 & -0.03 & -0.02 & 0. \\
 - & - & - & - & - & - & - & - & - & - & - & 1 & 0.54 & -0.64 & -0.08 & -0.02 & 0.05 & -0.01 & 0. & 0. \\
 - & - & - & - & - & - & - & - & - & - & - & - & 1 & 0.07 & -0.46 & 0.05 & 0.09 & -0.03 & -0.02 & 0. \\
 - & - & - & - & - & - & - & - & - & - & - & - & - & 1 & -0.01 & 0.1 & 0.03 & -0.02 & -0.01 & 0. \\
 - & - & - & - & - & - & - & - & - & - & - & - & - & - & 1 & 0.04 & 0.05 & -0.02 & -0.01 & 0. \\
 - & - & - & - & - & - & - & - & - & - & - & - & - & - & - & 1 & 0.32 & -0.11 & -0.15 & 0. \\
 - & - & - & - & - & - & - & - & - & - & - & - & - & - & - & - & 1 & -0.17 & -0.14 & 0. \\
 - & - & - & - & - & - & - & - & - & - & - & - & - & - & - & - & - & 1 & 0.9 & 0. \\
 - & - & - & - & - & - & - & - & - & - & - & - & - & - & - & - & - & - & 1 & 0. \\
 - & - & - & - & - & - & - & - & - & - & - & - & - & - & - & - & - & - & - & 1 \\
\end{array}
\right). \nn 
\eeq 
\end{tiny}

Despite appearances, the $W$ mass corrections $\delta m_w$ is not completely uncorrelated with the vertex correction, but the correlation coefficients are of order $0.001$, and are approximated as zero above. 

The results in \eref{resultsFinal} and \eref{correlationFinal} are enough to reproduce the full Gaussian likelihood function for our parameters.\footnote{This likelihood function is also available on request as a {\tt Mathematica} notebook.} 
Using the map in \eref{DEF_dgMap}, it can be translated into a likelihood for the Wilson coefficients in the Warsaw basis at the scale $\mu = m_Z$. 
In this step, the map should be used for $V\to {\bf I}$, since our fit results are formally valid in this limit. 
That can be evolved to other RG scales using e.g. the public code 
{\tt DsixTools}~\cite{Celis:2017hod} or {\tt Wilson}~\cite{Aebischer:2018bkb}. 
Our general likelihood can always be restricted to more constrained flavor scenarios~\cite{Aoude:2020dwv,Faroughy:2020ina} or to the universal scenario with oblique parameters~\cite{Wells:2015uba}.

\bibliographystyle{apsrev4-1}
\bibliography{references}

\end{document}